\RequirePackage{lineno}
\documentclass[linenumbers,author-year,preprint,showkeys,preprintnumbers,amsmath,amssymb,superscriptaddress]{revtex4}
\usepackage{graphicx}
\usepackage{graphics}
\usepackage{subfigure}
\usepackage{natbib}
\usepackage{dcolumn}
\usepackage{bm}
\usepackage{txfonts}
\usepackage{enumerate}
\usepackage{amsmath,empheq,mathrsfs}

\begin{document}


\centerline{\Large Highly coordinated nationwide massive travel restrictions are central} 
\centerline{\Large to effective mitigation and control of COVID-19 outbreaks in China}	

{\small {\vskip 12pt \centerline{Xingru Chen$^{1,2}$, \& Feng Fu$^{2,3}$}

\begin{center}
$^1$School of Sciences, Beijing University of Posts and Telecommunications, Beijing 100876, China\\
$^2$ Department of Mathematics, Dartmouth College, Hanover, NH 03755, USA\\
$^3$ Department of Biomedical Data Science, Geisel School of Medicine at Dartmouth, Lebanon, NH 03756, USA
\end{center}
}}

\vskip 30pt

\begin{minipage}{142mm}
\begin{flushleft}

{\textbf{Running title:}}\, China's COVID-19 lockdown 
{\textbf{Manuscript classification:}}\, Applied Math -- Network Epidemiology, Public Health -- Infectious Diseases Modeling\\
{\textbf{Author contributions:}}\, X.C., \& F.F. conceived the model, performed analyses, and wrote the manuscript. \\
{\textbf{Corresponding author:}} \\
Xingru Chen \\
\small {School of Sciences \\
Beijing University of Posts and Telecommunications \\
Beijing 100876, China \\
Email: xingrucz@gmail.com \\
Tel: +86 (010) 62281800 \\
Fax: +86 (010) 62281800 \\}
\end{flushleft}
\end{minipage}

\clearpage

\begin{center}
{
\begin{minipage}{142mm}
{\bf Abstract:} \, 

The COVID-19, the disease caused by the novel coronavirus 2019 (SARS-CoV-2), has caused graving woes across the globe since first reported in the epicenter Wuhan, Hubei, China, December 2019. The spread of COVID-19 in China has been successfully curtailed by massive travel restrictions that put more than 900 million people housebound for more than two months since the lockdown of Wuhan on 23 January 2020 when other provinces in China followed suit. Here, we assess the impact of China's massive lockdowns and travel restrictions reflected by the changes in mobility patterns before and during the lockdown period. We quantify the synchrony of mobility patterns across provinces and within provinces. Using these mobility data, we calibrate movement flow between provinces in combination with an epidemiological compartment model to quantify the effectiveness of lockdowns and reductions in disease transmission. Our analysis demonstrates that the onset and phase of local community transmission in other provinces depends on the cumulative population outflow received from the epicenter Hubei. As such, infections can propagate further into other interconnected places both near and far, thereby necessitating synchronous lockdowns. Moreover, our data-driven modeling analysis shows that lockdowns and consequently reduced mobility lag a certain time to elicit an actual impact on slowing down the spreading and ultimately putting the epidemic under check. In spite of the vastly heterogeneous demographics and epidemiological characteristics across China, mobility data shows that massive travel restrictions have been applied consistently via a top-down approach along with high levels of compliance from the bottom up. Our results show that such highly coordinated nationwide lockdowns have been able to effectively suppress the average number of secondary infections and thus central to mitigate and control \emph{early-stage} outbreaks and avert a massive health crisis otherwise. Our work sheds light on that large-scale coordination of collective actions, albeit costly in the short term such as complying with lockdown orders, is urgently needed to fight this pandemic and get our life back to normal.

\vspace*{0.5\baselineskip}

\end{minipage}
}
\end{center}

\clearpage


\section{Introduction}

Global health and humanity has been constantly threatened by emerging novel zoonotic diseases~\cite{lloyd2017predictions, royce2020mathematically}, such as Zika~\cite{petersen2016zika}, Ebola~\cite{leroy2005fruit}, and more recently the COVID-19~\cite{zhu2020novel, wu2020estimating}. The relentless siege of SARS-CoV-2, the pathogen causing COVID-19 infections~\cite{andersen2020proximal, munster2020novel}, has upended everyone's normal life and caused health crises, lockdowns, and economic percussions at an unprecedented pace and scale. The world has resorted to mandatory non-pharmaceutical interventions (NPI), including lockdowns, face-covering, and social distancing, so as to mitigate disease impact before effective pharmaceutical interventions (e.g., vaccines) become available~\cite{grein2020compassionate, hellewell2020feasibility, ferretti2020quantifying}. Such top-down approaches consider the society as a whole and attempt to optimize intervention measures from the perspective of central planners. On the other hand, adopting personal intervention measures such as complying with lockdown measures incurs a cost to oneself, but collectively protects the community especially these vulnerable. To address these issues, infectious disease dynamics has been an important research area in relevant mathematical and biological fields~\cite{anderson1992infectious, hethcote2000mathematics, levin1999population}. Over the years, researchers have proposed behavioral epidemiology as a means of integrating the study of epidemiology with an understanding of health decisions made by individual actors responding to infection risks~\cite{salathe2008effect, bauch2013social, bloom2014addressing, wadman2017vaccine, fu2017dueling, chen2019imperfect}.



Models of spatial epidemiology have been extensively studied using mathematical approaches combined with real data, with a focus on revealing the spatio-temporal pattern of epidemic spreading~\cite{bartlett1957measles, bolker1995space, elliott2004spatial}. In particular, it is shown that the persistence and resurgence of local community transmissions can be driven by movement between interconnected populations. To understand the persistence and cycles of measles outbreaks, prior work has found that the ubiquitous community structure and their intrinsic heterogeneity can hamper public health efforts to control and prevent childhood diseases~\cite{bolker1995space}. Moreover, previous studies take into account the network topology of communities along with their interconnected mobility in metapopulation models~\cite{colizza2007reaction, colizza2008epidemic}. These prior results provide novel insights into understanding the impact of individual movement on disease dynamics and implications for interventions. In recent years, with the increasing availability and particularly unprecedented dataset about human mobility data~\cite{song2010modelling, simini2012universal, csaji2013exploring}, it becomes feasible to study infectious disease and spatial epidemiology with more realistic considerations~\cite{balcan2009multiscale,wesolowski2012quantifying}.   

Of particular interest, previous research has demonstrated the effectiveness of travel restrictions to mitigate respiratory virus transmission, yet there are also significant limitations of this approach~\cite{mateus2014effectiveness, camitz2006effect, chong2012modeling, lin2020conceptual}. As aforementioned, epidemics in interconnected regions, partly due to the heterogeneity of underlying epidemiological characteristics, can exhibit complicated dynamics during an outbreak~\cite{bartlett1957measles, bolker1995space, colizza2008epidemic}. Among others, one of the important driving factors is movements, or more generally mobility patterns, that accounts for local commutes and domestic and international travels~\cite{balcan2009multiscale, wesolowski2012quantifying}. It is known that COVID-19 has a relatively long incubation period~\cite{li2020early} and can be contagious before the onset of symptoms (even asymptomatic transmission is more pervasive than previously thought~\cite{bi2020epidemiology}). Thus, in the very early phase of COVID-19 outbreaks when infrastructures like digital contact-tracing, high throughput testing capacity, and isolation stations are still lacking or insufficient to handle overwhelming exponential outgrowth of the epidemic, the somewhat brute-force lockdowns (through travel restrictions) seem to be the last resort to mitigate the disease impact and save time for development and deployment of other alternative interventions such as vaccines and treatments. However, as time goes by, testing, contact tracing, and isolation may provide a feasible approach for controlling local outbreaks of COVID-19~\cite{hellewell2020feasibility, ferretti2020quantifying}.

In the wake of COVID-19 outbreaks, a wealth of studies focus on investigating how reductions in international travels through various forms of lockdowns and travel bans would lessen the impact of the pandemic around the world~\cite{chinazzi2020effect, aleta2020data, pan2020association, zhang2020changes, lai2020effect}. Retrospectively, these studies invariably confirm the importance of curbing imported cases so as to prevent cross seeding of infections due to the spatial and temporal heterogeneity in epidemics, particularly when local transmissions are outweighed by the risk of imported cases~\cite{chinazzi2020effect}. Noteworthy,~\cite{jia2020population} analyzes how the cumulative population outflow from the epicenter impacts the timing of onset of local community transmissions in other receiving locations. The authors find a positive correlation between the two quantities, thereby enabling a leading time for predicting the onset of local outbreaks.


As an initial pandemic response when there are not any other pharmaceutical interventions at disposal, drastic lockdown imposes tremendous short-term cost but it can bring long-term positive impact if implemented with ideal timing and coordination. From the top-down management perspective, prior research has extensively investigated optimization of intervention policy using simple epidemic models~\cite{sethi1978optimal, wallinga2010optimizing, maharaj2012controlling, bolzoni2017time, huberts2020optimal, morris2021optimal}, including vaccination~\cite{abakuks1974optimal} and isolation~\cite{abakuks1973optimal,wickwire1975optimal}. In regard to adaptive social distancing, prior work uses evolutionary game theory~\cite{glaubitz2020oscillatory} or differential games~\cite{reluga2010game} from the perspective of individuals to understand factors of behavioral compliance. While these prior models shed light on adaptive social distancing, it remains urgently needed to examine the effectiveness of lockdowns and assess their actual impact on disease mitigation and control using a data-driven approach. For this purpose, changes in mobility patterns can be a good proxy to study the impact of non-pharmaceutical interventions (NPI) such as lockdowns and social distancing on behavior changes that help reduce community transmissions in an interconnected setting of spatial epidemiology.


In the meantime, it is not uncommon that some people protest against measures of lockdowns and NPIs~\cite{neumayer2021protest}, together with debates about their cost-effectiveness and impacts on health and society~\cite{born2020lockdowns, brodeur2021covid} (see a recent comprehensive review in~\cite{perra2021non}). To examine and validate conditions for lockdowns to be effective, we take a second look at China's containment efforts. Despite its vast size and huge populations, China's lockdowns have been able to mitigate and control local outbreaks through massive travel restrictions. In early 2020 when knowledge about the COVID-19 is still rather limited, the decision to lockdown came after many deliberations and amid tremendous uncertainty. Even so, to put a population of more than 1.4 billion on lockdowns and in some provinces strict home quarantines for an exceedingly long period is a hard, top-down decision that comes at an astronomical economic cost. Yet it turns out to have long-term positive impacts mainly because the stringent lockdowns have been implemented effectively.  

Here, we use a data-driven modeling framework to assess the effect of lockdowns on transmission reductions and improve our understanding of the necessity of uniform and highly synchronous lockdowns in light of the spatio-temporal pattern of COVID-19 outbreaks. To this end, we explore China's COVID-19 lockdowns as a concrete example, examine the level of synchrony of implementing travel restrictions across China, and quantify the impact of lockdowns on people movements. 

Our data analysis based on the massive mobility data reveals that lockdowns are implemented highly synchronously and uniformly at multi-levels, that is, between provinces (30 administrative regions) and within provinces (at the level of prefecture within a given province such as Hubei). As demonstrated in theoretical research about spatial epidemiology~\cite{bartlett1957measles, bolker1995space, colizza2008epidemic}, highly coordinated nationwide massive travel restrictions are central to effective mitigation and control of COVID-19 in China, especially during the early stage of epidemic outbreaks. Such massive travel restrictions ultimately lead to successful control of COVID-19, saving the country from the blink of a huge health crisis. Albeit the immeasurable loss of lives and economy, it is a remarkable achievement, which is made possible by the sacrifice of each and every one of ordinary people just like everyone elsewhere who has been impacted by this pandemic.

\section{Results}

In late December 2019, the outbreak of COVID-19 is first reported in Wuhan, Hubei, which is a central transportation hub (especially for trains). The situation is rapidly escalated to a public health emergency after local case surges and excessive hospitalizations that cause hospital overflow (Fig.~S1 in the SI). On 23 January 2020, the Chinese government imposed the largest scale of lockdown measure in human history amidst the busiest period of domestic travels around the Lunar New Year (Fig.~1) (as a reference, approximately 2.97 billion trips in total were made during a similar time window in the year before~\cite{chunyun18}). 

The implementation of lockdown is well coordinated across the nation with the highest level of epidemic response (Level 1). As shown in Fig.~1, both the influx and the outflux of travelers for each province approach rock bottom after a short period of chaos and panic. In some provinces, people desperately try to get in and get out amid the announcement of lockdowns and travel restrictions. Even worse, people may find themselves suddenly trapped in the middle of their way but still need to continue their trip for their final destinations. Such disruptions are reflected in the temporary increases in travel volume (indicated by the Baidu mobility index). For people who attempted last-minute moving in and moving out, inter-province mobility is not immediately suppressed but rather surges across many parts of the regions. Noteworthy, such initial lags in achieving actual mobility reductions are also attributed to the seasonal peaks in domestic travels near the Lunar New Year when people (including migrant workers) travel back home to reunite with their family (`chunyun', the biggest annual migration of humans in the world).

Due to the strict implementation of national lockdowns, the travel volume eventually approaches desired control target (Fig.~1). In particular, the epicenter Hubei experiences the strictest ever travel restrictions and there are barely any free people movements except for essential travels for an exceedingly long period (Fig.~1). As a matter of fact, Wuhan residents are strictly confined in their homes for months (in total 76 days) until early April 2020. In late February 2020, a month after the lockdown, many provinces lower their response levels and lift their lockdowns. As a consequence, travels rebound but still, the impact of lockdowns on mobility is long-lasting. Despite lower numbers of active cases, the reopening up efforts by the government see little effect as people need time to feel comfortable about traveling again due to the fear of lockdowns and the potential risk of infections. 

Fig.~2 shows the pairwise mobility index between provinces; the lump sum of each column and each row, respectively, gives the total volume of `move in' and `move out' in Fig.~1. The order of listed provinces is ranked according to their final epidemic size by the end of our study. The level of inter-province mobility appears to be correlated with their epidemic sizes, forming a cluster on the upper right corner (Fig.~2). 

The heatmap plots provide us with an intuitive visual guide for understanding the degree of interconnectedness between provinces in terms of their bidirectional travel volume (Fig.~2). After two weeks since the lockdown (Feb 4, 2020), the travel reduces to a bare minimum that is required to maintain essential living and work. The entire nation is paused at a massive scale and at a highly coordinated pace (Fig.~2). In most places, the intra-province mobility is reduced dramatically more than 90\%, and even more, the inter-province travels are cut almost at 100\%, in particular for the travel from and to the epicenter and other most affected provinces (see the upper rows, those provinces that had the largest outbreaks, of the heatmaps in the middle row of Fig.~2). Taken together, these results demonstrate that China's lockdowns are highly synchronous and effectively stop long-range spatial spreading due to domestic travels. 

As an outbreak unfolds, its emerging spatio-temporal pattern is highly dependent on the underlying multi-scale and multi-layer population structure, among others, most crucially on the mobility pattern~\cite{jia2020population, colizza2007reaction, bolker1995space, wesolowski2012quantifying}. People make local and non-local movements during which inevitable close contact/proximity with others can seed infections near and far. Fig.~3 presents an overview of the emergent spatio-temporal pattern of the COVID-19 outbreak in China. In accordance with Fig.~2, most affected provinces suffering the largest outbreaks by 10 March 2020 are those with the greatest levels of interconnectedness with, and thus receiving the largest population outflow from, the epicenter Hubei, including Guangdong, Henan, Zhejiang, Hunan, and Anhui (highlighted in Fig.~3a). Fig.~3b further demonstrates how the phase and magnitude of outbreaks in each province correlate with the cumulative population outflow received from Hubei. The greater outflow received from the epicenter, the earlier onset of the local outbreaks will be detected along with a larger number of cases (Fig.~3b). These data-based results provide a direct and intuitive rationale for synchronous lockdowns that are required to ultimately control and possibly eliminate infections, at least in the early phase of an epidemic when limited options other than costly non-pharmaceutical interventions are possible. Altogether, our analysis based on China's COVID-19 dataset suggests such population outflow from epicenters determines the timing and scale of the outbreaks (Fig.~3). 

To further quantify the synchrony of lockdowns implemented across China, we perform comprehensive comparative statistics of inter-province and intra-province time series of mobility data (Fig.~4). The correlations on the level of provinces range from $0.77$ to $0.98$ (with value one suggesting perfect synchrony) (Fig.~4a). Further zooming in, we take a close look at the prefecture level within the epicenter Hubei province: the synchrony of mobility changes due to well-coordinated lockdowns between the capital Wuhan and other cities within Hubei is significant and with little variations (correlations ranging from 0.85 to 0.95) (Fig. 4b). 

We further use an SEIR compartment model with a data-driven approach to infer the time-dependent transmission rate $\beta_i(t)$ (see Methods \& Model, and also the SI). This parameter reflects how well mobility patterns (reductions and changes) translate into effective transmission rates (infections via contacts) (Fig.~4c). We can see that the impact of lockdowns on reducing transmissions does not occur immediately, but lags a certain time (on average two weeks or so) to reach desired effective behavioral changes that eventually lead to reductions in transmissions. This result implies that lockdown measures need to last sufficiently long enough so as to see their positive mitigation impact, partly because people need some time to fully adjust to, and more importantly comply with, quarantine orders, especially strict home quarantine. 

We also estimate the effective basic reproductive ratio, $R_t$, in order to characterize the impact of interventions on controlling the epidemic over time. Fig.~5 shows the best estimated $R_t$ for each province, most of their values varying from 2 to 10. Owing to the unique demographics of each province, the heterogeneity of $R_t$ requires a distinct level of interventions. For example, the epicenter Hubei has an $R_t \sim 4$ prior to lockdowns, whereas its strongly interconnected province Guangdong has an $R_t \sim 15$. Despite such drastically different epidemiological characteristics and population densities and sizes, the universal lockdowns implemented almost synchronously have managed to contain the epidemic outbreaks in each province, which would have become too overwhelming to handle otherwise. Fig.~5 also reveals the intrinsic difference in the persistence of COVID-19 and the effectiveness of interventions across the nation. In Zhejiang and Shanghai (which are economically developed regions), the interventions are highly efficient and bring down the $R_t$ below one within days. In contrast, as for the epicenter Hubei, it takes a month to curb the level of infections below the critical threshold. Although lockdown is not a one-size-fits-all approach, there should be no question about its effectiveness, as long as implemented in synchrony across the target population, in control and mitigation of an emerging epidemic.


\begin{figure}
   \centering
   \includegraphics[width=0.8\columnwidth]{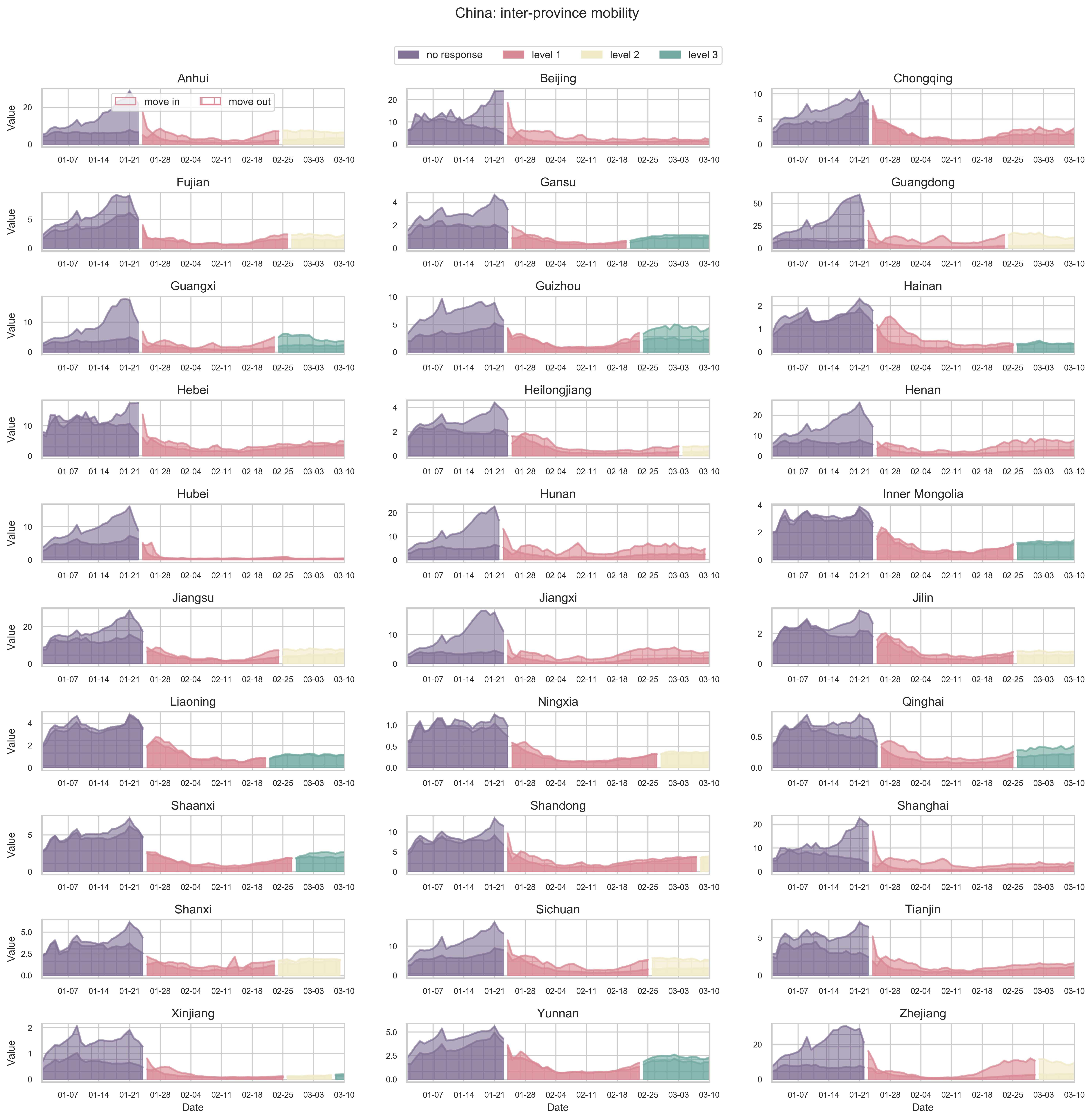} 
      \caption{Universal lockdowns across China along with a highly coordinated nationwide epidemic response. The plots each show the inter-province mobility, measured by daily total influx and outflux of inter-province travels using Baidu migration data, changes for the period from late December 2019 to March 2020. The Chinese government imposed by far the largest scale of strict travel restrictions on more than 11 million people (beyond) on January 23, 2020 (Level 1 response), amid the busiest period of the year for domestic travels (`chunyun', travels made during the Lunar New Year). Such massive travel restrictions have caused a dramatic reduction in travel volume, not only for the outflow from the epicenter Wuhan (Hubei) but also nationwide. Depending on the level of regional disease mitigation efforts, only a few provinces relax their travel restrictions (lowering from Level 1 to Level 3) a month later. The color corresponds to the level of response prior to and after the epidemic outbreak in Wuhan.}
   \label{fig1}
\end{figure}


\begin{figure}
   \centering
   \includegraphics[width=\columnwidth]{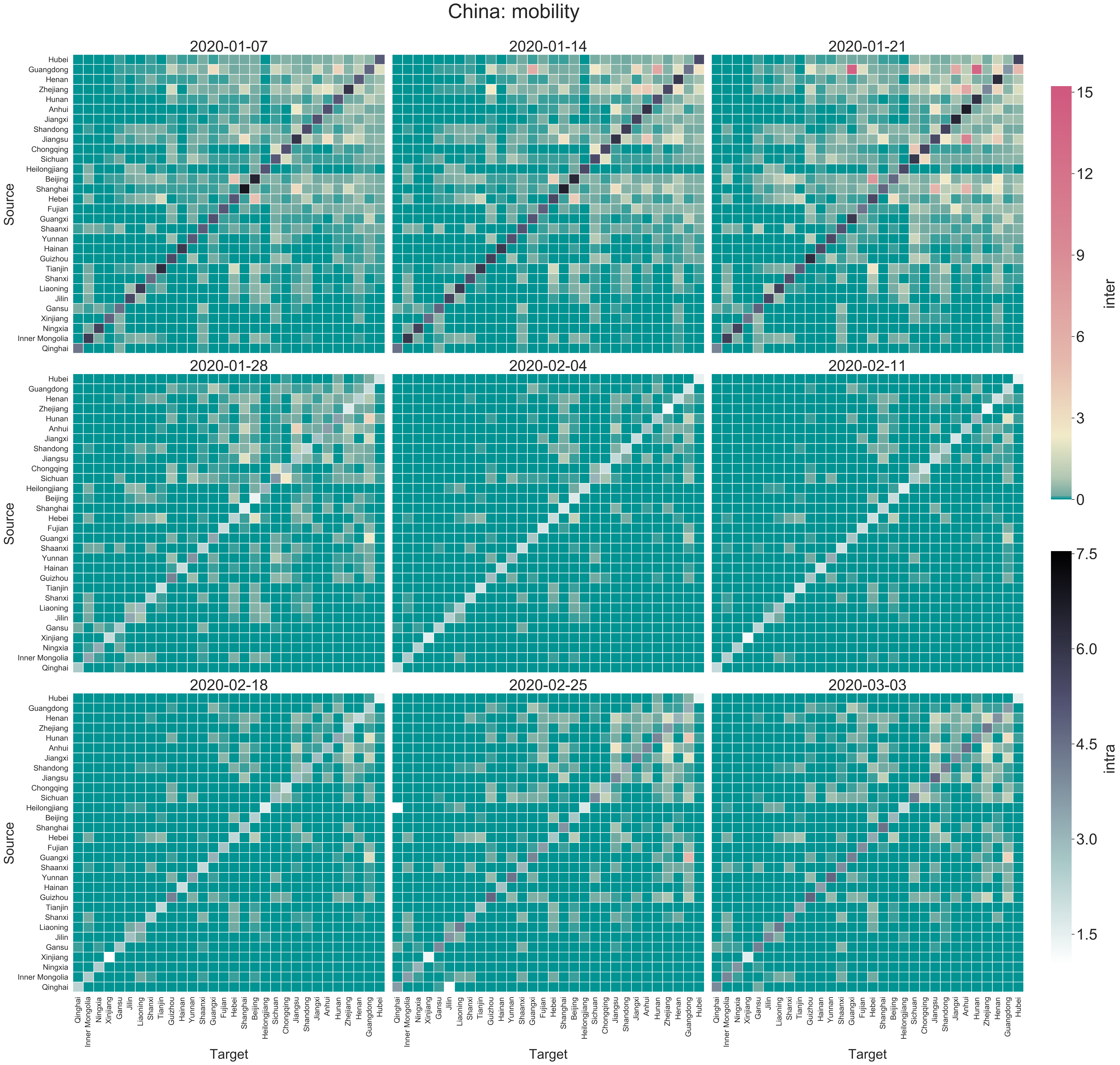} 
      \caption{Changes in inter-province and intra-province mobility over key dates throughout the epidemic outbreak. The non-diagonal elements of each heatmap plot show the migration index (a quantity proportional to the overall volume, as defined by Baidu) of pairwise travel destinations from province A (source) to B (target) while the diagonal the intra-province mobility index (travels within a given province). Prior to lockdowns, the travel peaks correspond to popular domestic travel routes during the Lunar New Year such as from Guangdong to Hunan (e.g., migrant workers return from coastal areas to inner lands to reunite with family). Both the inflow to and the outflow from Hubei (epicenter) are kept at extremely low levels except for essential travels that support epidemic response and basic living needs. These heatmaps complement Fig.~1 by providing more detailed views of mobility during the outbreak.}
   \label{fig2}
\end{figure}


\begin{figure}
   \centering
   \includegraphics[width=\columnwidth]{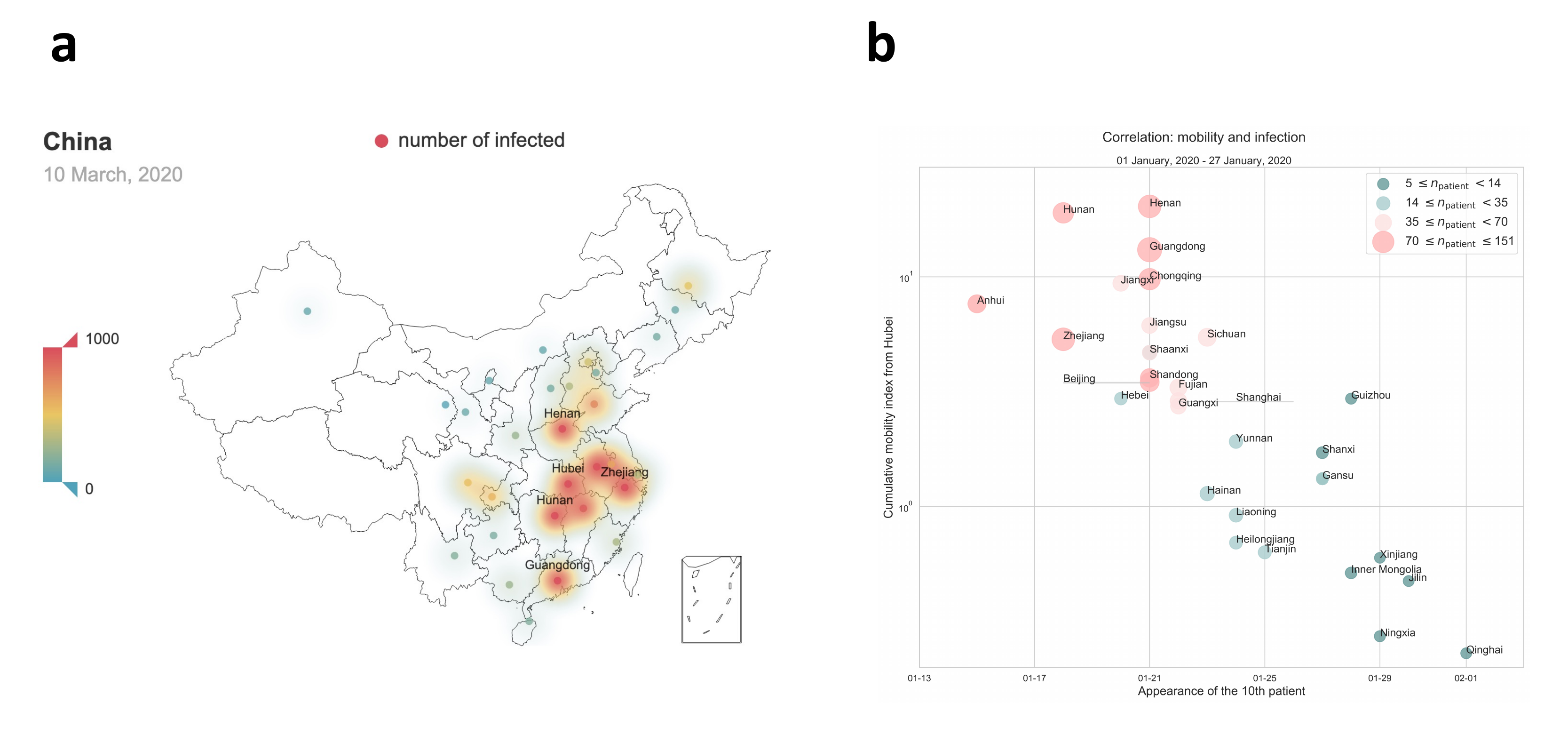} 
      \caption{Spatio-temporal pattern of early-stage epidemic spreading of COVID-19 in China. The phase and magnitude of local outbreaks within each province depend on the cumulative population inflow received from epicenter Hubei. Panel (a) shows the cumulative cases on the date 10 March 2020 and the red color corresponds to the most affected provinces. Panel (b) shows the timing of emerging infections (the appearance of the 10th diagnosed case) of each province versus the cumulative mobility index (proportional to the overall volume of travels received from Hubei for the period from 1 Jan 2020 to 27 Jan 2020). The dot size is proportional to the cumulative number of cases during that time window. Most affected provinces are those receiving greater population outflow from Hubei along with much earlier phases of epidemic outgrowth.
      }
   \label{fig3}
\end{figure}


\begin{figure}
   \centering
   \includegraphics[width=\columnwidth]{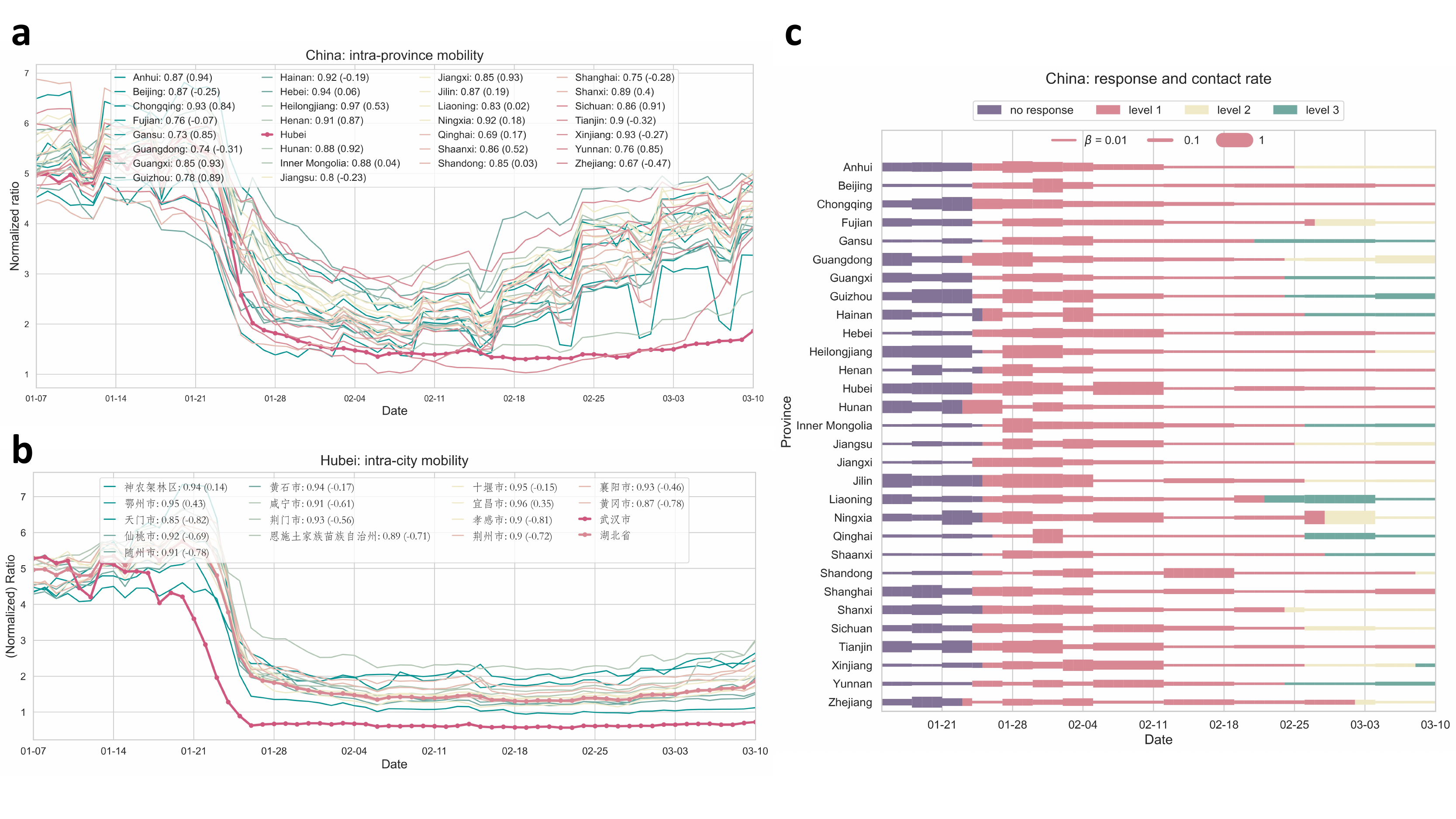} 
      \caption{Quantifying synchrony in reduced mobility due to national lockdowns and massive travel restrictions and assessing their impacts through reductions in disease transmissions inferred from our data-driven modeling. Panel (a) show the intra-province mobility and their strong correlations with the curve of Hubei province. Compared to the year before (numbers given in brackets in the legend), the mobility patterns exhibit significantly higher correlations, suggesting a high level of synchrony during the lockdown period. Panel (b) shows the correlation of mobility index among prefectures within Hubei province. Panel (c) shows the inferred transmission rates using a data-driven multi-compartment framework. In all provinces, reduced mobility levels translate to drastically suppressed transmissions. The effect of lockdowns on transmission reductions has seen a pronounced delay (varying by one or two weeks) for two reasons: (1) people need time to adjust to reduced social contacts despite decreasing mobility (2) local community transmissions cannot be easily controlled unless strict `cordon sanitaire' (home quarantine) is enforced. The color corresponds to the level of epidemic response.}
   \label{fig4}
\end{figure}

%
\begin{figure}
   \centering
   \includegraphics[width=\columnwidth]{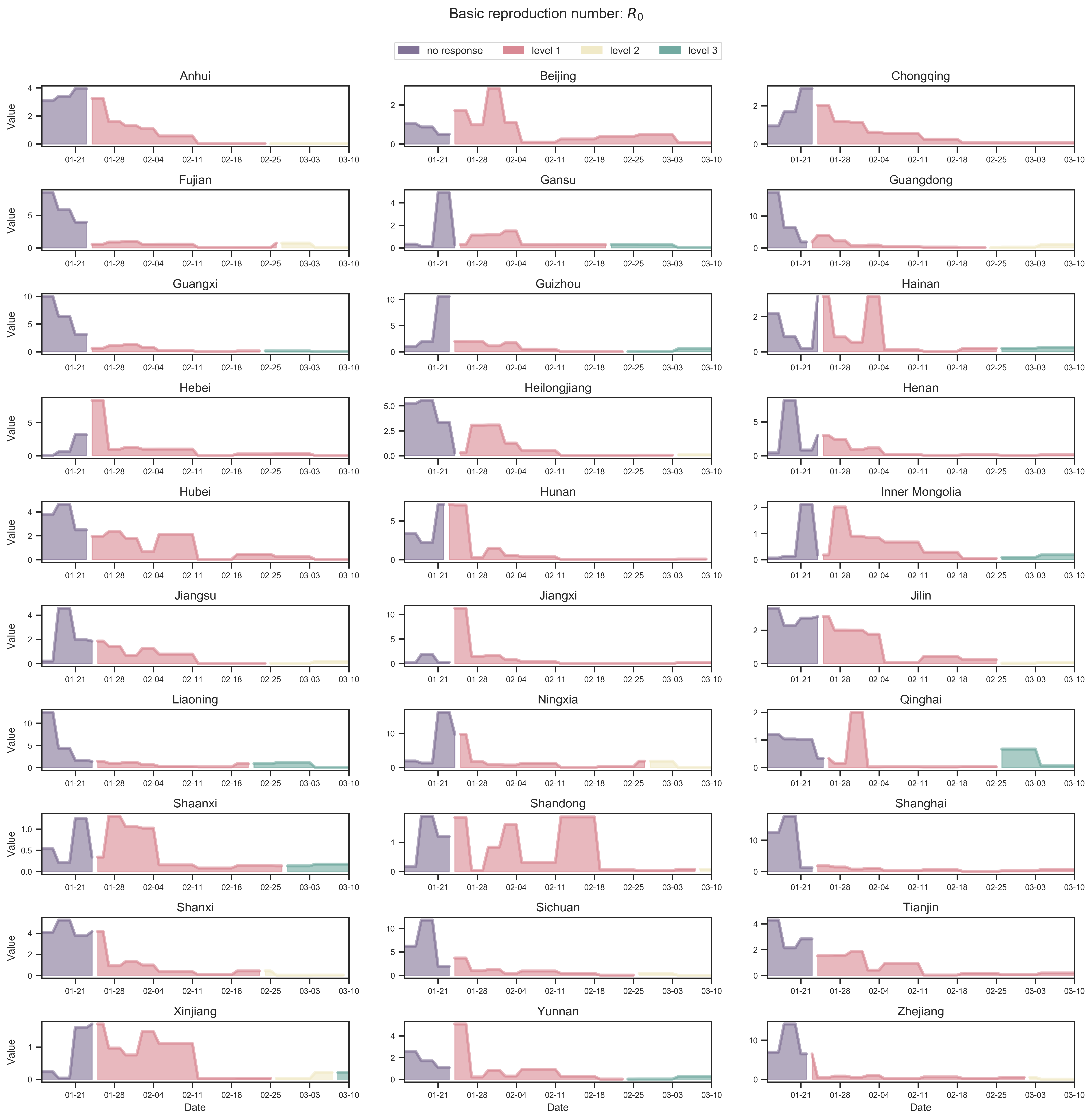} 
      \caption{Province-specific effective basic reproductive ratio, $R_t$, inferred from data-driven modeling. Highly coordinated nationwide massive travel restrictions are able to suppress infections across China, despite each province's distinct pace and magnitude of epidemic impact mitigation. The plot shows that province-specific $R_t$ is heterogeneous and has a distinctive pattern with respect to the implementation of local lockdown measures across provinces, but $R_t$ is uniformly suppressed after two weeks of nationwide lockdown and invariably drops well below one after one month.}
   \label{fig5}
\end{figure}

\section{Discussion \& Conclusion}

In the very early stage of an unprecedented outbreak of COVID-19 started in the epicenter, Wuhan, Hubei province, China, the Chinese government imposed by far the largest scale of strict travel restrictions on more than 11 million people (beyond) on January 23, 2020, amid the busiest period of the year for domestic travels (`chunyun', travels made during the Lunar New Year). Such massive travel restrictions have caused a dramatic reduction in travel volume, not only for the outflow from Wuhan (Hubei) but also nationwide (Figs.~1 and~2). Control measures like this help reduce the number of imported cases to other provinces, which can possibly slow down the onset of epidemic outbreaks in other regions and potentially weaken the impact of the disease. In this work, we use a data-driven approach to estimate the effectiveness of such massive travel restrictions in the mitigation of disease impact. Our work shows that highly coordinated massive travel restrictions are central to effective mitigation and control of COVID-19 outbreaks in China.

Pandemics are not new to human societies, yet tremendous challenges still remain particularly in the wake of the ongoing novel coronavirus pandemic~\cite{bolzoni2017time, maharaj2012controlling, kruse2020optimal}. Successful top-down management of the pandemic and governance of the collective in the face of infectious disease threats relies heavily on individual behavior and attitude changes from the bottom up~\cite{NatureHB2020Ed}. However, the \emph{tragedy of the commons} can arise as a result of `free-riding' in this important context~\cite{hardin1968tragedy}. Individuals may not follow disease intervention measures suggested by public health officials, especially if the epidemic curve is being bent down, but the uptick of cases, in turn, causes individuals to become more vigilant and increase their compliance levels. The feedback loop of this sort gives rise to oscillatory dynamics of disease prevalence and behavioral compliance to top-down public health measures, as seen in the current pandemic with multiple waves of infection~\cite{glaubitz2020oscillatory}. In addition to the social dilemma aspect of disease control, tremendous uncertainty associated with early detection of local community transmission and overall pandemic forecasting makes top-down scenario planning and optimization of intervention and mitigation extremely challenging~\cite{lipsitch2020defining}. As such, top-down and bottom-up modeling approaches need to go hand in hand in order to better inform public health efforts for effective disease interventions. 

One potential incomplete data issue of the present work is the aggregate Baidu mobility data based on phones we used. Admittedly, such mobility data can be underestimated due to population heterogeneity in phone usage and user privacy settings. However, as compared to a study published decade ago~\cite{fong2009digital}, China is now one of the leading countries in smartphone usage and ownership for both rural and urban populations: a recent Pew survey reported extraordinarily high smartphone ownership~\cite{poushter2017china}. Keep in mind our analysis is based on the levels of inter-province and intra-province mobility which already has a large population size. While accurate calibration for such potential sampling biases in mobility data is out of the scope of the present work, our results on understanding the spatio-temporal pattern of COVID-19 spreading and quantifying the impact of lockdowns are still of relevance and interest even as a case study of massive travel restrictions.
 
It remains an open problem to promote bottom-up behavior and attitude changes for the greater good~\cite{van2020using}. Compliance with public health recommendations and orders is an outstanding issue plaguing many parts of the world and greatly compromising efforts to mitigate the pandemic. Lockdowns had been attempted across the world, yet with drastically different outcomes. Many regions tried varying degrees of enforcement but met with resistance due to privacy and civil rights concerns. In contrast, China has a unique regime and governance structure to enforce a national lockdown through a well-considered top-down approach. 

Our work provides data-driven evidence for supporting highly coordinated lockdowns that should be implemented in the early onset of pandemics in order to be effective. Only when applied in concert with all regions are strict lockdowns effective. Across the world, other countries like Singapore and New Zealand have also seen successes in containing COVID-19 using well-coordinated national lockdowns. Undoubtedly, such lockdowns come at a huge cost --- business shutdowns, worker layoffs, and lack of child care, just to name a few -- and inflict economic repercussions. On the other hand, it is necessary to do so in the early phase of the global pandemic when effective pharmaceutical interventions (like vaccines and anti-viral treatments) are lacking or still under rapid development. 

With the increasing options of interventions and especially ramping up vaccination, it becomes hopeful to get our life back to normal. Massive travel restrictions are no longer needed to contain case surges~\cite{karatayev2020local}, provided that responsive and targeted local lockdowns by means of high-precision contact tracing as well as testing and isolation are in place. Since the full-scale national lockdowns in January 2020, local outbreaks due to imported cases from time to time in China have been successfully controlled and eliminated using such prompt, targeted testing and isolation so as to avert serious and costly national lockdowns repeatedly.

In conclusion, the pandemic has fundamentally shaped the whole world and reminds us of the importance of pandemic preparedness and global health management. The worst of all are discrimination and hate crimes around the globe~\cite{reny2020xenophobia}. Large-scale cooperation is urgently needed to solve many challenging issues facing our common humanity. Fighting this pandemic is yet another wake-up call for that.

\section{Methods \& Model}

\paragraph{\textbf{Model description}.} 

Our modeling framework builds on multi-scale behavioral epidemiological spreading processes that incorporate mobility patterns~\cite{chinazzi2020effect, wesolowski2012quantifying, balcan2009multiscale} of inter-province migrations (which affect the spatial spreading among provinces). Specifically, we consider a Susceptible-Exposed-Infected-Recovered (SEIR) model in a metapopulation structure with migration. The systems of ODEs describe the dynamics in continuous time $t$, that is, days since the disease outbreak:

\begin{eqnarray*}
\frac{dS_i(t)}{dt} & = & -\beta_i(t) S_i(t) \frac{I_i(t)}{N_i(t)} - \sum_{j, j\neq i}\alpha_{ij}(t)S_i(t) + \sum_{j, j\neq i} \alpha_{ji}(t)S_j(t);\\
\frac{dE_i(t)}{dt} & = & \beta_i(t) S_i(t)  \frac{I_i(t)}{N_i(t)} - \sigma_i(t) E_i(t) - \sum_{j, j\neq i}\alpha_{ij}(t)E_i(t) + \sum_{j, j\neq i} \alpha_{ji}(t)E_j(t);\\
\frac{dI_i(t)}{dt} & = & \sigma_i(t) E_i(t) - \gamma_i(t) I_i(t);\\
\frac{dR_i(t)}{dt} & = & \gamma_i(t) I_i(t).
\end{eqnarray*}

Here, the subscript $i$ refers to the $i$th compartment on the provincial level. $N_i(t) = S_i(t) + E_i(t) + I_i(t) + R_i(t)$ is the total population size of compartment $i$ at time $t$. $\sum_{j, j\neq i}\alpha_{ij}(t) \left[S_i(t) + E_i(t) \right]$ is the total outflow from compartment $i$ to other compartments, and $\sum_{j, j\neq i} \alpha_{ji}(t) \left[S_i(t) + E_i(t) \right]$ is the total inflow to compartment $i$ from other compartments. To parameterize migration flows between compartments, we use the real provincial level mobility data from the Baidu Map service, which provides aggregate level tracking of domestic travels on a daily basis. The impact of lockdowns and travel restrictions on aggregate behavioral responses/changes that lead to transmission reductions is characterized by the province-specific transmission rate $\beta_i(t)$. More detailed modeling and data analysis can be found in the SI.

Epidemiological model parameters:
\begin{itemize}
\item
unit of time $t$: day.
\item
$\beta_i(t)$: transmissibility rate, which can be time dependent, due to lockdown efforts (home quarantine and travel restrictions)~\cite{verity2020estimates}.
\item
$1/\sigma_i(t)$: incubation period, which ranges from $1\sim 14$ days~\cite{li2020early,bi2020epidemiology}.
\item
$1/\gamma_i(t)$: number of days remain infectious, which ranges from $1\sim 14$ days~\cite{he2020temporal}.
\item
$R_0 = \beta_i(0)/\gamma_i(t)$: the initial values of $R_0$ are bounded within $1.4 \sim 4$~\cite{abbott2020temporal,sanche2020novel}.
\end{itemize}

\noindent{\textbf{Uncertainty quantification and sensitive analysis.}} We apply the \textbf{dual annealing} algorithm to perform a nonlinear least square fitting procedure for estimating time-varying epidemiological parameters, denoted by a vector $\hat{\theta}$, combined with real mobility data. We further calculate the covariance matrix $\text{cov}(\hat{\theta}) = s^2(F'F)^{-1}, \text{with}\, F = \left. \frac{\partial f(\theta)}{\partial \theta}\right|_{\theta = \hat{\theta}}$, and hence standard errors of the estimated parameters (the diagonal elements of $\hat{\theta}$). The covariance matrix contains complete information about the uncertainty of parameter estimations. Following a Student's $t$-distribution, the confidence interval at $(1 - 2\alpha)$ significance is given by $\hat{\theta}_{1 - 2\alpha} = \hat{\theta} \pm t_{n - p}^{\alpha}\sqrt{\text{diag}\,\text{cov}(\hat{
\theta})}$ (see, e.g., the uncertainty quantification for our predicted curve in Fig. S11 in the SI, where the uncertainty of prediction, denoted by the shaded area, propagates as a function of the mean behavior of the spreading dynamics).

\paragraph{\textbf{Datasets}.} Mobility data is obtained from Baidu~\url{https://qianxi.baidu.com}. COVID-19 data is collected and curated by DXY and archived at~\url{https://github.com/BlankerL/DXY-COVID-19-Data}.

\paragraph{\textbf{Open code}.} Source code is available at the GitHub repository~\url{https://github.com/fudab/China-COVID-19-mobility}. An interactive website can be found at~\url{https://fudab.github.io/covid-19/china}.

\section*{Acknowledgements}
We thank Timmy Ma, Xin Wang, and Daniel Escudero for helpful discussions when initializing this work. F.F. is indebted to Dan Rockmore and Nicholas Christakis for stimulating discussions on pandemic modeling and data analysis. X.C. gratefully acknowledges the generous faculty startup fund support by BUPT. F.F. is grateful for support from the Bill \& Melinda Gates Foundation (award no. OPP1217336), the NIH COBRE Program (grant no.1P20GM130454), the Neukom CompX Faculty Grant, the Dartmouth Faculty Startup Fund, and the Walter \& Constance Burke Research Initiation Award.

\section*{References}

\end{document}